\documentclass[preprint]{aastex}
\usepackage{bm}
\usepackage{epsfig}
\usepackage{graphicx}
\usepackage{color}
\usepackage{subfigure}
\usepackage[colorlinks=true,linkcolor=blue,citecolor=blue]{hyperref}
\usepackage{natbib}

\shorttitle{Energy extent of nonthermal spectra}
\shortauthors{Werner et.~al.}

\begin{document}

\title{The extent of power-law energy spectra in collisionless relativistic magnetic reconnection in pair plasmas}

\author{G.~R.\ Werner, D.~A.\ Uzdensky} 
\affil{Center for Integrated Plasma Studies, Physics Department, \\390 UCB, 
University of Colorado, Boulder, CO 80309, USA}
\email{Greg.Werner@colorado.edu}
\and
\author{B.\ Cerutti\altaffilmark{1}}
\affil{Department of Astrophysical Sciences, Princeton University, Princeton, NJ 08544, USA}
\altaffiltext{1}{Lyman Spitzer Jr. Fellow} 
\and
\author{K.\ Nalewajko\altaffilmark{2}, M.~C.\ Begelman\altaffilmark{3}}
\affil{JILA, University of Colorado and National Institute of
Standards and Technology, \\440 UCB, Boulder, CO 80309-0440, USA}
\altaffiltext{2}{
NASA Einstein Postdoctoral Fellow (PF3-140130) at the
Kavli Institute for Particle Astrophysics and Cosmology, Stanford University, 
and Stanford Linear Accelerator Center, 2575 Sand Hill Rd, Menlo Park, CA 94025, USA
}
\and
\altaffiltext{3}{Department of Astrophysical and Planetary Sciences, 
391 UCB, Boulder, CO 80309, USA}


\begin{abstract}

Using two-dimensional particle-in-cell simulations, we characterize
the energy spectra of particles accelerated by
relativistic magnetic reconnection (without guide field) in
collisionless electron-positron plasmas, for a wide range of 
upstream magnetizations~$\sigma$ and system sizes~$L$.
The particle spectra are well-represented 
by a power law $\gamma^{-\alpha}$, with a combination of 
exponential and super-exponential high-energy cutoffs, proportional to $\sigma$ and $L$,
respectively. For large $L$ and $\sigma$, the power-law index $\alpha$ approaches about 1.2.

\end{abstract}


%

\maketitle


\section{Introduction}

Magnetic reconnection is a fundamental plasma physics process in which magnetic field rearrangement and relaxation rapidly converts magnetic energy into particle energy~\citep{Zweibel_Yamada-2009}.  Reconnection is believed to drive many explosive phenomena in the universe, from Earth magnetospheric substorms and solar flares to high-energy X-ray and $\gamma$-ray flares in various astrophysical objects. 
Quite often, the radiation spectra of these flares, and hence the energy distributions of the emitting particles, are observed to be non-thermal (e.g., characterized by power laws).
Therefore, understanding the mechanisms of {\it nonthermal particle acceleration} and determining the observable characteristics---such as the power-law index and high-energy cutoff---of the resulting particle distribution, is an outstanding problem in modern heliospheric physics and plasma astrophysics.

Of particular interest in high-energy astrophysics is the role of \emph{relativistic reconnection}---which occurs when the energy density of the reconnecting magnetic field, $B_0^2/8\pi$, exceeds the rest-mass energy density $n_b m c^2$ of the ambient plasma, leading to relativistic bulk outflows and plasma heating to relativistic temperatures---as a potentially important mechanism for nonthermal particle acceleration to ultra-relativistic energies (with Lorentz factors $\gamma \gg 1$) in various astrophysical sources~\citep{Hoshino_Lyubarsky-2012}.  In particular, this process has been invoked to explain energy dissipation and radiation production in electron-positron (pair) plasmas over multiple scales in pulsar systems---e.g., in the pulsar magnetosphere near the light cylinder, in the striped pulsar wind, and in the pulsar wind nebula (PWN)  \citep{Lyubarsky:1996,Lyubarsky:2001, Coroniti:1990, Uzdensky_etal-2011, Cerutti_etal-2012a, Cerutti_etal-2013, Cerutti_etal-2014a,Cerutti_etal-2014b, Sironi_Spitkovsky-2011, Uzdensky:2014}.  In addition, relativistic reconnection in pair and/or electron-ion plasmas is believed to play an important role in gamma-ray bursts (GRBs) \citep{Drenkhahn:2002,Giannios:2007,McKinney:2012} and in coronae and jets of accreting black holes, including  AGN/blazar jets, e.g., in the context of TeV blazar flares \citep{Giannios:2009,Nalewajko:2011}.

\begin{sloppypar} 
Nonthermal particle acceleration is essentially a kinetic
(i.e., non-fluid) phenomenon.  Although fluid simulations with
test particles have been used to study particle acceleration,
particle-in-cell (PIC) simulations include kinetic effects self-consistently.
A number of PIC studies have investigated 
particle acceleration in collisionless relativistic pair-plasma reconnection \citep{Zenitani_Hoshino-2001, Zenitani_Hoshino-2005, Zenitani_Hoshino-2007, Zenitani_Hoshino-2008, Jaroschek_etal-2004a, Lyubarsky_Liverts-2008,  Liu_etal-2011, Sironi_Spitkovsky-2011, Bessho_Bhattacharjee-2012, Kagan_etal-2013, Cerutti_etal-2012b, Cerutti_etal-2013, Cerutti_etal-2014a, Cerutti_etal-2014b, Liu2015scaling,Kagan2015relativistic}; the best evidence for nonthermal particle distributions was provided recently by~\citet{Sironi_Spitkovsky-2014,Guo:2014}.  
Whereas previous studies have identified {\it power-law slopes} of nonthermal spectra, the important question of the energy {\it extent} of these power laws has not been systematically addressed. 
\end{sloppypar}

In this Letter we present a comprehensive two-dimensional (2D) PIC investigation of non-thermal particle acceleration in collisionless relativistic reconnection in a pair plasma without guide magnetic field.  
In particular, we characterize the dependence of the resulting energy distribution function on the system size~$L$ and the upstream ``cold'' magnetization parameter $\sigma \equiv B_0^2/(4\pi n_b m_e c^2)$ (relativistic reconnection requires $\sigma \gg 1$).
We find empirically that relativistic reconnection produces a high-energy spectrum that is well represented by a power law with exponential and super-exponential cutoffs:\begin{eqnarray} \label{eq:dist}
  f(\gamma) = \frac{dN}{d \gamma} & \propto & \gamma^{-\alpha} 
                  \exp \left(-\gamma/\gamma_{c1}
                             -\gamma^2/\gamma_{c2}^2 \right)
.\end{eqnarray}
The different cutoffs serendipitously allow us to distinguish different scalings with $\sigma$ and $L$:
$\gamma_{c1} \sim 4 \sigma$ depends on $\sigma$, while 
$\gamma_{c2} \sim 0.1 L/\rho_0$ depends on $L$ (here $\rho_0 \equiv m_e c^2/eB_0$ is the nominal Larmor radius).

Equality of the two cutoffs, $\gamma_{c1} \simeq \gamma_{c2}$,
defines a critical size $L_c \simeq 40 \sigma \rho_0$
separating the small- and large-system regimes.
We find that for large systems ($L/\sigma\rho_0 \gg 40$),
the energy spectrum of accelerated particles (hence $\gamma_{c1}$)
is essentially independent of~$L$.
Importantly (as we discuss later), $L_c$ is approximately the
length at which a current layer, with thickness equal to the average
Larmor radius $\rho_e = \bar{\gamma}\rho_0$, becomes tearing-unstable
and breaks up into multiple plasmoids and secondary current sheets.
[Here, $\bar{\gamma} m_e c^2$ is the average dissipated energy per
background particle, 
$\bar{\gamma} \simeq \kappa (B_0^2/8\pi) / (n_b m_e c^2) = \kappa \sigma/2$;
in our simulations $\kappa \simeq 0.6$, so 
$\bar{\gamma} \simeq 0.3 \, \sigma$.]
Therefore, we propose that (at least in 2D with an initially cold background plasma) reconnection in the 
plasmoid-dominated regime yields a high-energy particle spectrum that is
predominantly independent of system size $L\gg L_c$.  Consequently, 
nonthermal particle acceleration in {\it huge}, 
astrophysically-relevant systems may be studied via merely 
{\it large} simulations, i.e., with $L\gtrsim L_c$.


\section{Simulations}

This study focuses on reconnection in 2D without guide field ($B_z=0$).   Although some important differences in the reconnection dynamics  emerge between 2D and 3D, such as the development of the drift-kink instability \citep{Zenitani_Hoshino-2008}, 
the dimensionality is not believed to affect the particle energy spectra at late stages \citep{Sironi_Spitkovsky-2014,Guo:2014,Daughton:2014pc,Drake:2014pc}. 
Working in 2D (much less costly than 3D) enabled investigation of large system sizes.


We simulate systems of size $L_x=L_y=L$ with periodic boundary conditions and two antiparallel reconnection layers. The two layers begin as relativistic Harris current sheets \citep{Kirk_Skjaeraasen-2003} {with upstream magnetic field $B_x=B_0$ and} a peak drifting 
plasma simulation-frame-density $n_d$ (at the layer centers) that is 10 times the uniform background density~$n_b$.  
A small (1\%) initial magnetic-flux perturbation facilitates reconnection onset.
Electrons and positrons in each Harris layer drift (in opposite directions) with average velocity $\beta_d c = 0.6c$, and initial Maxwell-J\"{u}ttner temperature $\theta_d \equiv k_B T_d/m_e c^2 = \sigma/16$;
the initial layer half-thickness is 
$\delta = (8/3) \theta_d \, \rho_0 = \sigma \rho_0/6$.
The background plasma is initially at rest, with temperature $\theta_b \ll \sigma$; however, 
due to the finite grid instability \citep{Birdsall1980plasma}, the background plasma is expected to 
heat until its Debye length is resolved, which occurs at a temperature 
of order $\theta_D \sim \sigma/512$ for $\Delta x=\sigma \rho_0/32$.

The simulations begin with $N_p = 128$ (macro)particles per grid cell with cell sizes $\Delta x = \Delta y  = \theta_d \,\rho_0 / 2 = \sigma \rho_0/32 \approx 0.2 \delta$
(except for $\sigma=3$, where the mildly-relativistic particles allowed $\Delta x = \theta_d \rho_0$ without sacrificing accuracy). 
The total energy is conserved within 1\% during each
simulation.
Convergence tests with respect to $\Delta x$ and $N_p$ indicate that our simulations are well resolved and, in particular, that the high-energy parts of the particle distributions are robust.  

The Vorpal code \citep{Nieter:2004}, employed for this study, uses an explicit electromagnetic PIC time advance, with Yee 
electromagnetics 
and a relativistic Boris particle push.

To determine the power-law index $\alpha$ and the energy cutoffs $\gamma_{c1}$ and $\gamma_{c2}$ as functions of the upstream magnetization $\sigma$ and the system size~$L$,
we ran simulations with $\sigma$ = 3, 10, 30, 100, 300, 1000, and, for each~$\sigma$, a range of system sizes up to $L/\sigma\rho_0 = 100$ for $\sigma=1000, 300$, up to $L/\sigma\rho_0=200$ for $\sigma=100, 30, 10$, and up to $L/\sigma\rho_0=400$ for~$\sigma=3$.


\section{Results}  
\begin{figure}[tbp]
\setlength{\tabcolsep}{1pt}
\begin{tabular}{ll}
\includegraphics*[width=8.cm,trim=4mm 0 4mm 0]{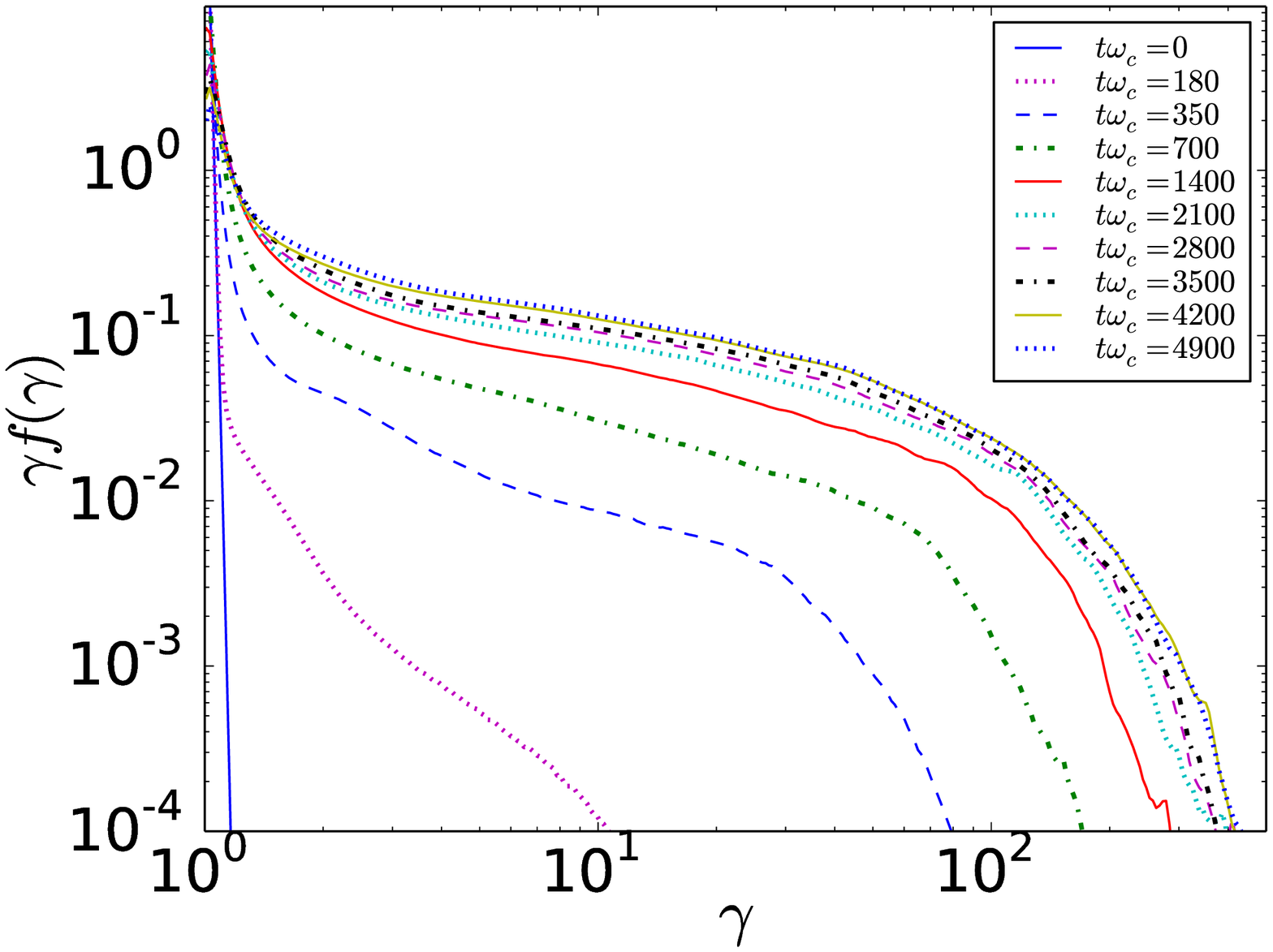}%
&\includegraphics*[width=8.cm,trim=3mm 0 4mm 0]{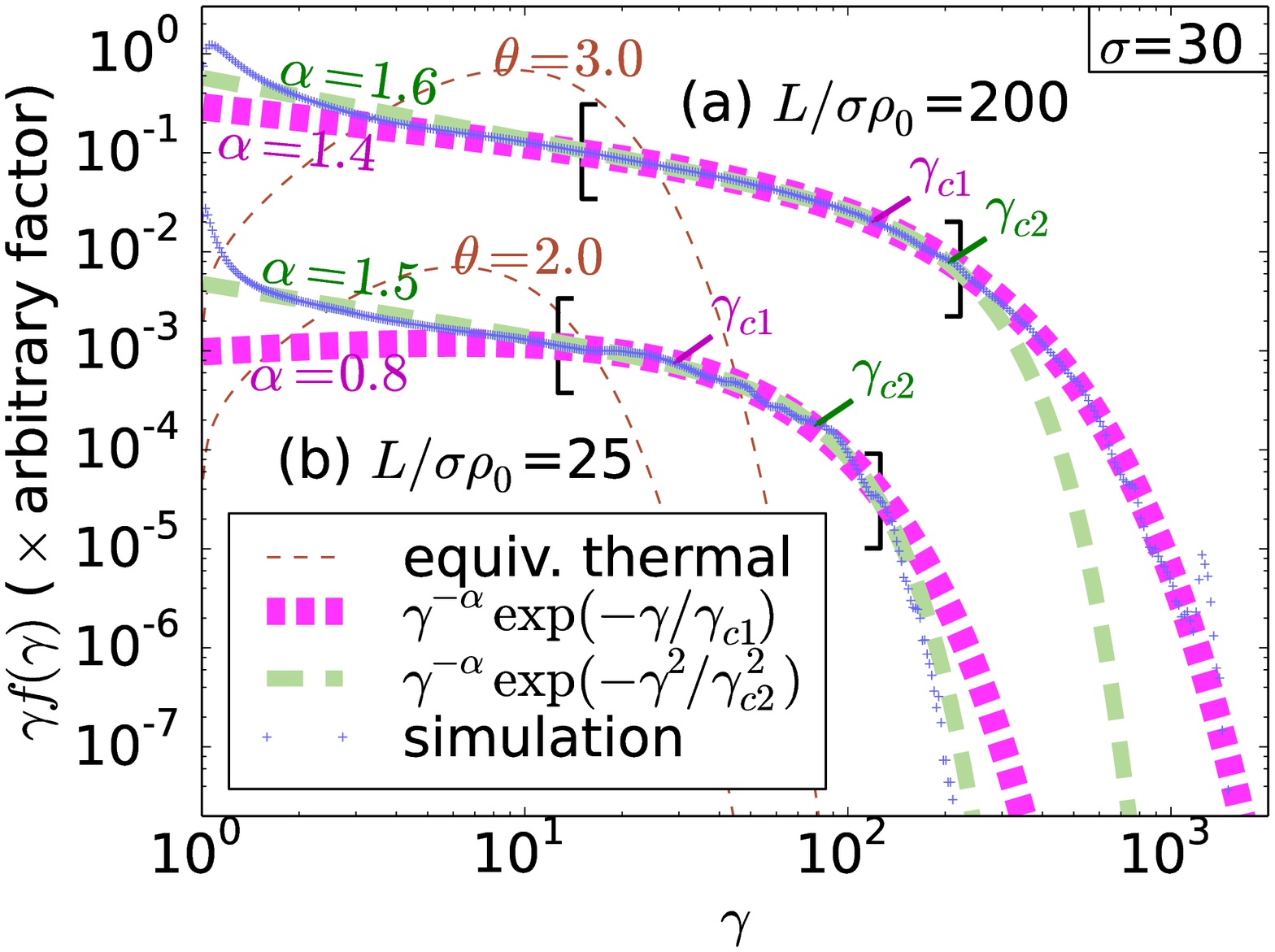}
\end{tabular}
\caption{(Left) Time evolution of the particle energy spectrum for a run with $\sigma=30$ and $L/\sigma \rho_0=200$.  
Reconnection ceases at
$t\omega_c\approx 4300$, but 
the shape of the high-energy spectrum is the same for
$t\omega_c\gtrsim 2000$ [$\omega_c\equiv c/(\theta_d \rho_0)$].
(Right) An exponential cutoff (short dashes) fits the energy spectra
better for large-$L$ simulations (a), while a super-exponential cutoff (long dashes) fits better for
small $L$ (b). Brackets mark $[\gamma_{f1},\gamma_{f2}]$, 
where the displayed fits were performed.
Thin-dashed lines show Maxwell-J\"{u}ttner distributions with equivalent
total energies.  Considering many fits (e.g., with different $\gamma_{f1}$, $\gamma_{f2}$), we determined 
for (a) $\alpha \in [1.38,1.49]$, $\gamma_{c1} \in [119,157]$,
$\gamma_{c2}$ too large/uncertain to measure;
for (b) $\alpha \in [1.31,1.48]$, $\gamma_{c1}$ too large/uncertain to measure,
$\gamma_{c2}\in [39,44]$.
\label{fig:distVsTime}
\label{fig:fits}}
\end{figure}

We focus on the energy distribution of 
background particles, excluding the initially-drifting particles,
which contribute negligibly to the overall distribution for large $L$.  
Evolution to a nonthermal distribution
proceeds rapidly (Fig.~\ref{fig:distVsTime}); and the
shape of the high-energy spectrum,
as characterized by $\alpha$ and $\gamma_{c1/2}$, ceases
to evolve well before all available flux has reconnected,
especially for large systems.

We find that the late-time high-energy spectrum is a power law with a high-energy cutoff significantly above the average particle energy,
in agreement with~\citet{Sironi_Spitkovsky-2014}.
We further observe (Fig.~\ref{fig:fits}) that spectra for large systems
have exponential cutoffs, 
$\exp(-\gamma/\gamma_{c1})$,
while small systems have sharper cutoffs, which we empirically model with a super-exponential
$\exp(-\gamma^2/\gamma_{c2}^2)$.
We therefore fit all spectra with the universal form of Eq.~(\ref{eq:dist}) to determine the power-law index $\alpha$ and the cutoffs 
$\gamma_{c1}$,~$\gamma_{c2}$; for small systems, the best-fit $\gamma_{c1}$ is typically much larger than $\gamma_{c2}$ (hence irrelevant 
and highly uncertain), while for large $L$, $\gamma_{c2}$ is larger and uncertain.

Each spectrum is fit to Eq.~(\ref{eq:dist}) over
an interval $[\gamma_{f1}, \gamma_{f2}]$,
chosen as large as possible while maintaining a good fit.  
Because spectra depart from a power law at lowest energies, and because
of increased noise at highest energies, larger fitting intervals
yield unacceptably poor fits.
Noise is reduced (and fit improved) by
averaging over short time intervals and,
if available, over multiple simulations (identical except for 
randomized initial particle velocities).
Because the choices 
of acceptable fit quality and
the durations of averaging intervals are somewhat subjective,
we perform many fits using different choices, and
finally report the median values
with ``error'' bars encompassing the middle 68\% of the fits 
(i.e., $\pm 1$ standard
deviation if the data were Gaussian-distributed);
small error bars thus demonstrate insensitivity to the fitting process.
Very uncertain and large (hence irrelevant) cutoff values are discarded.

\begin{figure}[tbp]
\includegraphics[width=8.6cm]{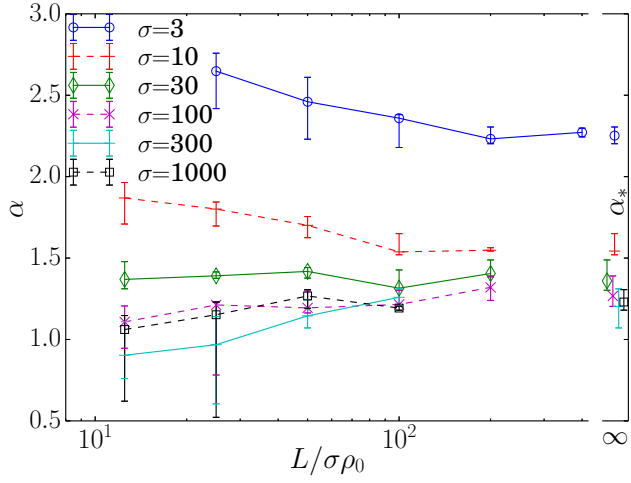}
\caption{Measured power-law indices $\alpha$ vs.
$L$, with extrapolations ($\alpha_*$) to $L\rightarrow \infty$ (cf. Fig.~\ref{fig:alphaVsSigma}).
\label{fig:alphaVsL}}
\end{figure}

\begin{figure}[tbp]
\includegraphics[width=8.6cm]{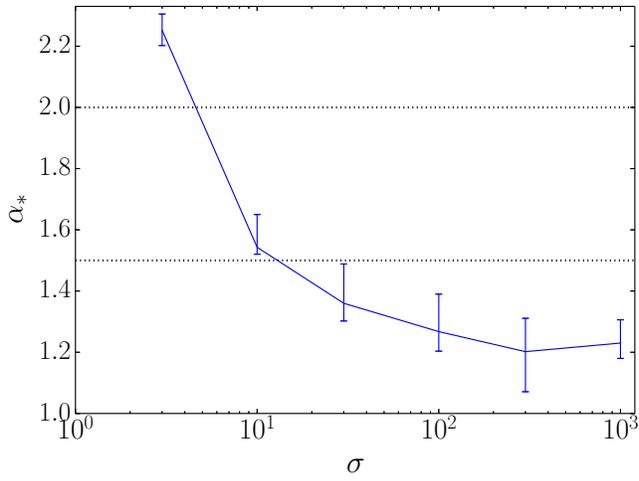}
\caption{Power-law index $\alpha_*$ vs. upstream magnetization $\sigma$.
\label{fig:alphaVsSigma}}
\end{figure}

By applying this fitting procedure to the background particle spectrum
for each different value of
$(L,\sigma)$, we mapped out $\alpha$, $\gamma_{c1}$, and $\gamma_{c2}$ 
as functions of $\sigma$ and $L$, up to sufficiently large $L$ to
estimate the asymptotic values 
$\alpha_*(\sigma) = \lim_{L\rightarrow \infty} \alpha(\sigma,L)$ 
(Fig.~\ref{fig:alphaVsL}).
We find that~$\alpha_*(\sigma)$ starts above 2
for modest~$\sigma$, and decreases to $\alpha_*(\sigma) \approx 1.2$
in the ultra-relativistic limit of $\sigma \gg 1$
(Fig.~\ref{fig:alphaVsSigma}), a result that is 
broadly consistent with previous studies 
\citep{Zenitani_Hoshino-2001,Jaroschek_etal-2004a,Lyubarsky_Liverts-2008,Sironi_Spitkovsky-2014,Guo:2014,Melzani:2014b}; while our measurement is closer to 1.2 than 1, the uncertainty is too large to
rule out $\alpha_* \rightarrow 1$, predicted by some \citep{Larrabee_etal-2003,Guo:2014}.

\begin{figure}[tbp]
\includegraphics[width=8.6cm]{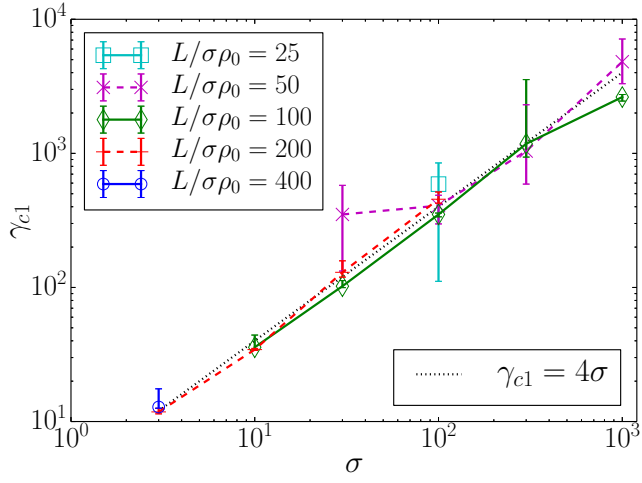}
\caption{The exponential cutoff $\gamma_{c1}$ scales linearly with
magnetization $\sigma$.
\label{fig:gc1VsSigma}}
\end{figure}

\begin{figure}[tbp]
\includegraphics[width=8.6cm]{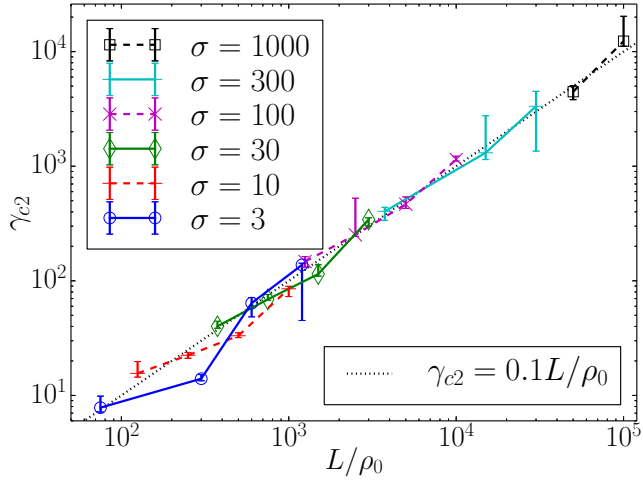}
\caption{The super-exponential cutoff $\gamma_{c2}$ scales linearly with
system size $L$.
\label{fig:gc2VsL}}
\end{figure}

In contrast to the power-law index $\alpha$, 
the energy \emph{extent} of the power law has received relatively little attention in relativistic reconnection literature \citep{Larrabee_etal-2003,Lyubarsky_Liverts-2008}.
We find that the high-energy cutoffs scale as
$\gamma_{c1} \sim 4 \sigma$ (\emph{independent of} $L$)
and 
$\gamma_{c2} \sim 0.1 L/\rho_0$ (\emph{independent of} $\sigma$)
(Figs.~\ref{fig:gc1VsSigma},~\ref{fig:gc2VsL}).
Thus $L/\sigma \rho_0 \ll 40$ implies $\gamma_{c2} \ll \gamma_{c1}$, and
a super-exponential cuts off the power-law at an energy determined by 
the system size.
Larger system sizes $L/\sigma \rho_0 \gg 40$ have $\gamma_{c1} \ll \gamma_{c2}$, and so $\gamma_{c1}$ determines where the power law ends,
independent of $L$.

\section{Discussion}

The scaling of the high-energy cutoffs can be explained in terms
of the distance a particle could travel within
the reconnection field $E_z\sim \beta_r B_0$ 
(where $\beta_r \sim 0.1$ is the reconnection rate).
By calculating analytic trajectories in fields 
\emph{around a single $X$-point},
Ref.~\citep{Larrabee_etal-2003} concluded that
$f(\gamma)\propto \gamma^{-1} \exp(-\gamma/\Gamma_0)$ 
with $\Gamma_0 = 12 e \beta_r^2 B_0 \ell_x / m_e c^2 \sim e E_z \ell_x/m_e c^2 \sim 0.1 \ell_x/\rho_0$, with
$\ell_x$ being the size of the reconnection region in $x$ \footnote[4]{
  The $x$-extent of the reconnection region is the relevant length
  here because the
  calculation considered motion in the $xz$-plane subject to
  fields uniform in $z$, so escape (hence cessation of acceleration) 
  was possible only through motion in $x$.
}, a result that was supported by 2D PIC simulation in \citet{Lyubarsky_Liverts-2008}.

In general, small systems reconnect mainly with one $X$-point, 
so $\ell_x\sim L$ and $\Gamma_0 \sim 0.1 L/\rho_0$,
which equals our $\gamma_{c2}$. 
(The observed super-exponential form presumably results from the
simulation's boundary conditions.)

In large systems, however, the tearing instability breaks up 
current layers with full-length greater than $\ell_{\rm tear} \sim 100 \bar{\delta}$, 
where $\bar{\delta}$ is the
layer half-thickness \citep{Loureiro:2005,Ji:2011}, resulting in a 
hierarchy of layers ending with {\it elementary} layers, which
are marginally stable against tearing 
\citep{Shibata:2001,Loureiro_etal-2007,Uzdensky:2010}.  
The half-thickness of elementary (single $X$-point, laminar) 
layers should be about the average Larmor radius
$\bar{\delta} \sim \rho_e = \bar{\gamma} \rho_0$ 
\citep{Kirk_Skjaeraasen-2003}.
Although \citet{Larrabee_etal-2003} considered single $X$-point
reconnection, we propose that their formula for $\Gamma_0$ 
can also be used in the context of plasmoid-dominated reconnection 
in large systems if applied
to elementary layers (instead of the entire global layer):
$\ell_x \sim \ell_{\rm tear} \sim 100 \bar{\gamma}\rho_0 \sim 30 \sigma \rho_0$ 
(instead of $\ell_x\sim L$).  Then, 
$\Gamma_0 \sim 0.1 \ell_{\rm tear}/\rho_0 \sim 3 \sigma $,
which is essentially our $\gamma_{c1}$ 
(and consistent with the measurement of
 $\Gamma_{0}= 35$ for $\sigma =9$ in \citet{Lyubarsky_Liverts-2008}).

\begin{sloppypar} 
This explanation of high-energy-cutoff scaling in terms of
elementary layer lengths may be 
robust despite the potentially important roles played by other acceleration
mechanisms \citep{Hoshino_Lyubarsky-2012}.
For example, significant additional acceleration may occur 
within contracting plasmoids 
\citep{Drake:2006,Dahlin:2014,Guo:2014,Guo:2015} or---especially for
the highest-energy particles---in the (anti-)reconnection electric field of
secondary plasmoid mergers \citep{Oka:2010,Sironi_Spitkovsky-2014,Nalewajko:2015}.
\end{sloppypar} 

It is interesting to compare our high-energy cutoffs to the upper bound imposed on a power-law distribution by a finite energy budget.
When $1<\alpha <2$, most of the kinetic energy resides in high-energy particles, so the available energy per particle $\bar{\gamma}\sim 0.3\sigma$ limits the extent of the power law.
If $f(\gamma)\sim \gamma^{-\alpha}$ extends from $\gamma_{\rm min}$ to $\gamma_{\rm max} \gg \gamma_{\rm min}$, then 
$\bar{\gamma} \approx [(\alpha-1)/(2-\alpha)] \gamma_{\rm min}^{\alpha-1}
\gamma_{\rm max}^{2-\alpha}$ \citep{Sironi_Spitkovsky-2014}.
For $\alpha \approx 1$, $\gamma_{\rm max}$ can extend well beyond $\bar{\gamma}$, but $\gamma_{\rm max}/\bar{\gamma}$ depends weakly on system parameters, consistent with our finding $\gamma_{c1} \sim \bar{\gamma} \sim \sigma$.  
E.g., for $\alpha = 1.2$,  $\gamma_{\rm max}/\bar{\gamma} \approx (10^3\, \bar{\gamma}/\gamma_{\rm min})^{1/4}$.
However, when $\alpha > 2$  (e.g., for low $\sigma$), the energy budget imposes no upper bound, since
$\int_{\gamma_{\rm min}}^\infty \gamma \gamma^{-\alpha} d\gamma $ is finite.  Nevertheless, for $\sigma = 3$ where $\alpha_* > 2$, we observe $\gamma_{c1} \sim 4\sigma$, the same as for smaller $\alpha_*$.

The exponential cutoff at energies above $\gamma_{c1} \sim 4\sigma \sim 10 \bar{\gamma}$ has important astrophysical implications for particle acceleration in systems such as pulsar magnetospheres, winds, PWN, and relativistic jets in GRBs and AGNs.
Our results (insofar as they are ultra-relativistic) can be generalized to
relativistically-hot upstream plasmas by scaling all the energies by $\bar{\gamma}_b$, the average Lorentz factor of background particles. 
The ``hot'' magnetization $\sigma^{(\textrm{hot})} \equiv B_0^2/(4\pi n w)$
therefore parameterizes similar simulations, since 
the relativistic specific enthalpy $w$ also scales with 
$\bar{\gamma}_b$ [i.e., $w=\bar{\gamma}_b m_e c^2 + p_b/n_b$, where $p_b$ is the background plasma pressure; for $\bar{\gamma}_b \gg 1 $, $w\approx (4/3) \bar{\gamma}_b m_e c^2$].\footnote[5]{
  Because the finite grid instability heats the background plasma
  until its Debye length is resolved \citep{Birdsall1980plasma}, 
  the resolution prevents us from
  obtaining values of $\sigma^{(\textrm{hot})}$ above a few 
  hundred.  For our simulations with $\sigma \lesssim 100$,   
  $\sigma^{(\textrm{hot})} \approx \sigma$; however, for 
  $\sigma \gtrsim 300$, the numerical heating reduces the value of 
  $\sigma^{(\textrm{hot})}$.
}
For example, our reconnection-based  model \citep{Uzdensky_etal-2011, Cerutti_etal-2012a, Cerutti_etal-2013, Cerutti_etal-2014a, Cerutti_etal-2014b} for high-energy $\gamma$-ray flares in the Crab PWN~\citep{Abdo:2011, Tavani:2011} relies upon acceleration of a
significant number of particles from $\bar{\gamma}_b\sim 3 \times 10^6$ to $\gamma\gtrsim 10^9$.  If, to achieve this, we need $\gamma_{c1} > 10^9$, then direct extrapolation of the results from this Letter would require
$\sigma^{(\textrm{hot})} \gtrsim (1/4) \gamma_{c1}/(w/m_ec^2) \approx 60$;
this should be comparable (via scaling equivalence) to simulations presented in this
work with $\sigma\sim 60$ (corresponding to a power-law index $\alpha_* \sim 1.3$).
This required $\sigma^{(\textrm{hot})}$ is significantly higher than what is expected in the Crab Nebula.  
However, here we analyzed the entire spectrum of background particles, while 
\citep{Cerutti_etal-2012b} suggested that bright flares observed
in the Crab Nebula result from preferential focusing of the highest-energy particles into tight beams with energy spectra that differ from the entire spectrum.
We also note that our present simulations are initialized with a Maxwellian plasma, whereas the ambient plasma filling the Crab Nebula has a power-law distribution, which may result in a higher high-energy cutoff.

\section{Conclusion} 
We ran a series of collisionless relativistic pair-plasma magnetic reconnection simulations with no guide field, covering a wide range of system sizes $L$ and upstream magnetizations $\sigma \geq 3$.  We observed
acceleration of the background plasma particles to a nonthermal energy distribution
$f(\gamma) \sim \gamma^{-\alpha(L,\sigma)}$ 
with a high-energy cutoff.  The cutoff energy is proportional to the maximum length of elementary, single $X$-point layers, which is limited by $L$ in small systems, and by the secondary tearing instability in 
large systems.
For small systems ($L \ll 40 \sigma \rho_0$) we observe
$ f(\gamma) \sim \gamma^{-\alpha} \exp(- \gamma^2/\gamma_{c2}^2)$ with
$\gamma_{c2}\sim 0.1 L/\rho_0$, and for large systems,
$ f(\gamma) \sim \gamma^{-\alpha} \exp(-\gamma/\gamma_{c1})$
with $\gamma_{c1} \sim 4 \sigma$.
As $L$ becomes large,
the power-law index $\alpha(L,\sigma)$ asymptotically approaches 
$\alpha_*(\sigma)$, which in turn decreases to $\approx 1.2$ as $\sigma \rightarrow \infty$.
This characterization of power-law slope and high-energy cutoffs can be used to link ambient plasma conditions (i.e., $\sigma$) with observed radiation from high-energy particles, to investigate the role that reconnection plays in high-energy particle acceleration in the universe.

\acknowledgments

This work was supported by DOE grants DE-SC0008409 and DE-SC0008655, NASA
grant NNX12AP17G, and NSF grant AST-1411879.
Numerical simulations were made possible by the
Extreme Science and Engineering Discovery Environment (XSEDE),
which is supported by National Science Foundation (NSF) grant number
ACI-1053575---and in particular by the NSF
under Grant numbers 0171134, 0933959, 1041709, and
1041710 and the University of Tennessee through the use of the 
Kraken
computing resource at the National Institute for Computational Sciences
(www.nics.tennessee.edu/).  This work also used the Janus supercomputer,
which is supported by the NSF (award number CNS-0821794) and the
University of Colorado Boulder; the Janus supercomputer is a joint effort
of the University of Colorado Boulder, the University of Colorado
Denver, and the National Center for Atmospheric Research.
We gratefully acknowledge the developers responsible for the Vorpal 
code.

\bibliographystyle{apj}

\end{document}